\documentclass[graybox]{svmult}

\usepackage{mathptmx}       
\usepackage{helvet}         
\usepackage{courier}        
\usepackage{type1cm}        
\usepackage{makeidx}         
\usepackage{graphicx}        
\usepackage{multicol}        
\usepackage[bottom]{footmisc}

\makeindex             

\begin{document}

\title*{Can the hint of $\delta_{CP}$  from T2K also indicate the hierarchy and octant ?}
\author{Monojit Ghosh, Srubabati Goswami and Sushant K. Raut}
\institute{Monojit Ghosh \at Physical Research Laboratory, Navrangpura, Ahmedabad 380009, India, \email{monojit@prl.res.in} \and
Srubabati Goswami \at Physical Research Laboratory, Navrangpura, Ahmedabad 380 009, India, \email{sruba@prl.res.in}
\and Sushant K. Raut \at Department of Theoretical Physics, School of Engineering Sciences, KTH Royal Institute of Technology - AlbaNova University Center,
Roslagstullsbacken 21, 106 91 Stockholm, Sweden, \email{raut@kth.se}}

\maketitle

\abstract{The T2K neutrino data has already given a hint for the best-fit value of the leptonic CP phase $\delta_{CP}$ as $-90^\circ$. In this 
paper we ask the question that if this hint is confirmed by the subsequent neutrino and anti-neutrino runs of T2K, then can it also give any information
about the other two remaining unknown oscillation parameters - the neutrino mass hierarchy and octant of $\theta_{23}$. We find that if T2K runs in only neutrino mode
with its full targeted exposure, then $\delta_{CP} = -90^\circ$ would indicate the true hierarchy as normal and the true octant as higher. On the other hand
if T2K runs in equal neutrino and anti-neutrino mode then the true hierarchy can be confirmed as normal but the octant will remain undetermined.
We have also studied the effect of anti-neutrino runs on CP sensitivity of T2K. 
}

\section{Introduction}
\label{sec:1}
At present neutrino oscillation physics stands at a very interesting juncture. Among the six oscillation parameters that describe neutrino
oscillation, experiments have measured the values of the three mixing angles (i.e., $\theta_{12}, \theta_{13}$ and $\theta_{23}$), and the
two mass squared differences (i.e., $\Delta_{21} = m_2^2 - m_1^2$ and $|\Delta_{31}| = |m_3^2 - m_1^2|$) with considerable precision. 
The remaining unknowns in this sector are :
(i) the neutrino mass hierarchy (i.e normal or NH: $m_3 > m_1$ or inverted or IH: $m_3 < m_1$), (ii) the octant of $\theta_{23}$ (lower or LO: 
$\theta_{23} < 45^\circ$ or higher or HO: $\theta_{23} < 45^\circ$) and (iii) the exact value of $\delta_{CP}$ \cite{global}. Recently there is a hint of $\delta_{CP}=-90^\circ$ 
driven mainly by the combination of T2K and reactor data \cite{t2krecent}. This hint comes from the T2K running in the neutrino mode with $8\%$ of the total exposure
(7.8 $\times$ $10^{21}$ protons on target viz. POT) \cite{t2krecent}.
In this article, we consider the possibility of determining neutrino mass hierarchy and octant of $\theta_{23}$ in future runs of T2K. 
We assume that the T2K hint of $\delta_{CP}=-90^\circ$
will be established by further neutrino and anti-neutrino runs. In that case we argue that if T2K completes its run in pure neutrino mode then the data would
also suggest the true hierarchy as NH and true octant as HO. This is because for the other combinations of hierarchy and octant, the true value of $\delta_{CP}=-90^\circ$
can also be mimicked by other wrong values of $\delta_{CP}$. Therefore a clear hint of $\delta_{CP}=-90^\circ$ would not be possible. On the other hand
if T2K completes its run in 50$\%$ neutrino and 50$\%$ anti-neutrino mode then the true hierarchy will be NH and no conclusion can be drawn about octant as in this case
both LO and HO can give a clear hint of  $\delta_{CP}=-90^\circ$. In this context we also study the role played by the anti-neutrinos in $\delta_{CP}$
measurements of T2K and identify the true combinations of hierarchy and octant where anti-neutrino runs can be useful. Our work is important in the sense that
an early hint of the major unknowns of neutrino oscillations will be useful in planning future facilities.

\section{Degeneracies in $P_{\mu e}$}
\label{sec:2}
The T2K is a long-baseline experiment in Japan having a baseline of 295 km. The expression of the appearance channel probability
relevant for this baseline is given by \cite{akhmedov}
\begin{eqnarray}
P_{\mu e }&=& 4 s_{13}^2 s_{23}^2 \frac{\sin^2(\hat{A}-1)\Delta }{(\hat{A}-1)^2}  \\  \nonumber
&&
+ 2 \alpha s_{13} \sin{2\theta_{12}} \sin{2\theta_{23}}\cos{(\Delta + \delta_{CP})} \times \\ \nonumber
&& \qquad\qquad \frac{\sin{\hat{A}\Delta}}{\hat{A}} \frac{ \sin{(\hat{A}-1)\Delta} }{\hat{A}-1}  + {\cal{O}}(\alpha^2) ~. 
\label{P-emu}
\end{eqnarray}
where 
$s_{ij} \equiv \sin{\theta_{ij}}$, $\alpha = \Delta_{21}/\Delta_{31}$,  
$\Delta = \Delta_{31}L/4E$, $L$ is the baseline and $E$ is the energy of the neutrino. 
$\hat{A} = 2\sqrt{2} E G_F n_e/\Delta_{31}$, is the matter term with Fermi constant $G_F$ and electron density $n_e$. 
The determination of $\delta_{CP}$ in T2K suffers due to the presence of degeneracies. This implies different sets of parameters giving
equally good fit to the data. In view of the large value $\theta_{13}$, at present two types of degeneracies are important : 
(i) Hierarchy-$\delta_{CP}$ degeneracy: $P_{\mu e}(\delta_{CP}, \Delta) = P_{\mu e}(\delta_{CP}^\prime, -\Delta^\prime)$ i.e., for a given octant,
probability for NH can be the same as probability for IH \cite{degen} and 
(ii) Octant-$\delta_{CP}$ degeneracy : $P_{\mu e}(\theta_{23}^{LO},\delta_{CP})
= P_{\mu e}(\theta_{23}^{HO},\delta_{CP}^\prime)$ i.e., for a given hierarchy, probability for LO can be the same as probability for HO \cite{lisi}.
The wrong hierarchy and/or wrong octant solutions can occur for a value of $\delta_{CP}$ different than the true value, which affects the CP
sensitivity as well.
For anti-neutrinos, the hierarchy-$\delta_{CP}$ degeneracy behaves in the same way as that of neutrinos \cite{novat2k}
but the nature of octant-$\delta_{CP}$ degeneracy is different in anti-neutrinos as compared to neutrinos \cite{uma}. This can be understood in the following way.
For neutrinos (anti-neutrinos), $P_{\mu e }$ is
higher for NH (IH) and lower for IH (NH). 
However the relative sign of $\delta_{CP}$ is also opposite for neutrino 
and anti-neutrino probabilities.
This causes the hierarchy-$\delta_{CP}$ degeneracy to behave in the similar fashion for 
both neutrinos and anti-neutrinos. On the other hand 
$P_{\mu e }$ is lower for LO and higher for HO  
for both neutrinos and anti-neutrinos and this makes the octant-$\delta_{CP}$
degeneracy to behave differently for neutrinos and anti-neutrinos.
This signifies that combination of neutrino and anti-neutrino
channel is important for removal of octant-$\delta_{CP}$ degeneracy 
but not for removal of hierarchy-$\delta_{CP}$ degeneracy.

\section{Results}
\label{sec:3}
We simulate T2K using the software GLoBES \cite{globes}. We consider a total T2K exposure of $8 \times 10^{21}$ POT.
In our analysis LO (HO) corresponds to $\theta_{23} = 39^\circ(51^\circ)$
and NH (IH) corresponds to $\Delta_{31} = +2.4(-2.4) \times 10^{-3}$. In our figures we have calculated $\chi^2$
using the Poisson formula and plotted in the y axis. 

\subsection{Hint for hierarchy and octant}
\label{subsec:3a}

\begin{figure*}[t]
  \hspace{-0.2in}
  \includegraphics[scale=0.5]{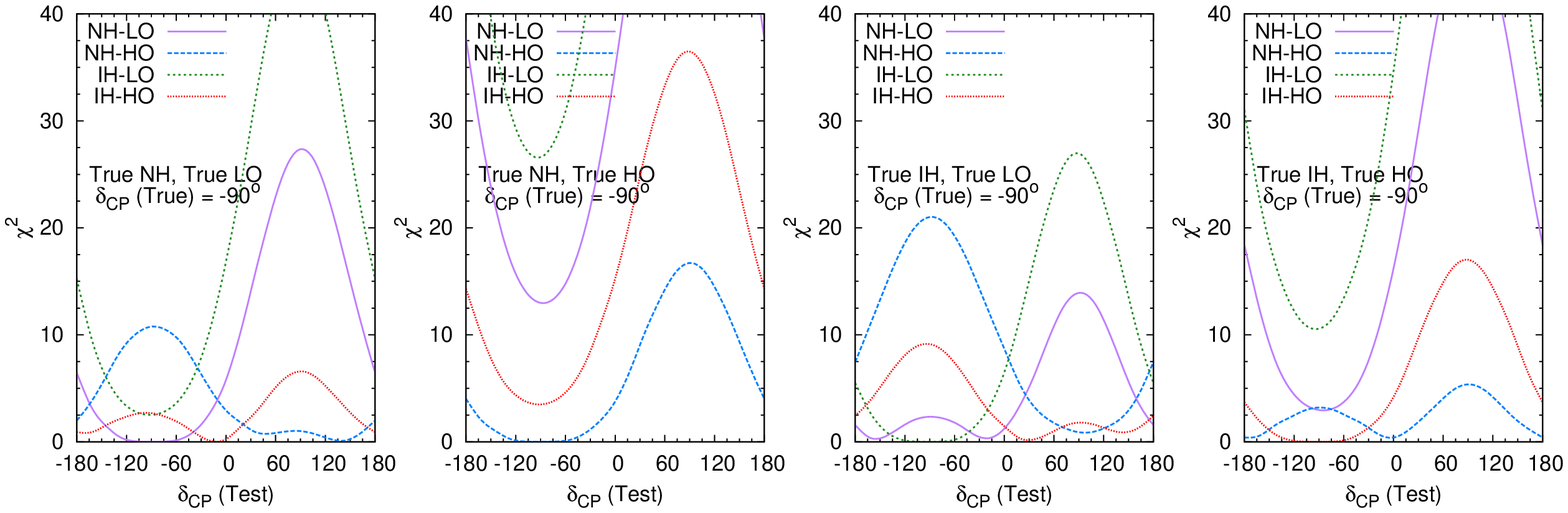}
  \caption{$\delta_{CP}$ sensitivity of T2K neutrino run}
  \label{cpnurun}
  \end{figure*}
  
 \begin{figure*}[t]
  \hspace{-0.2in}
  \includegraphics[scale=0.5]{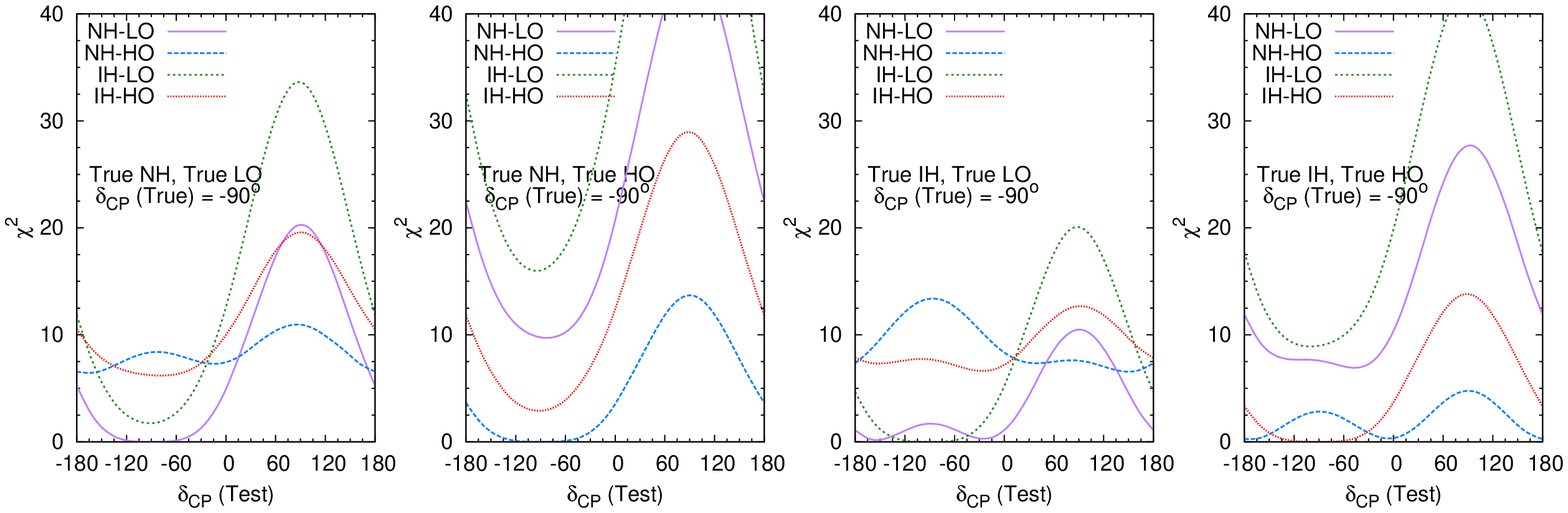}
  \caption{$\delta_{CP}$ sensitivity of T2K for equal neutrino+anti-neutrino run 
  (Total pot $= 8 \times 10^{21}$)}
  \label{cpantinurun}
  \end{figure*}
  
   \begin{figure*}
   \hspace{-0.2in}
   \includegraphics[scale=0.5]{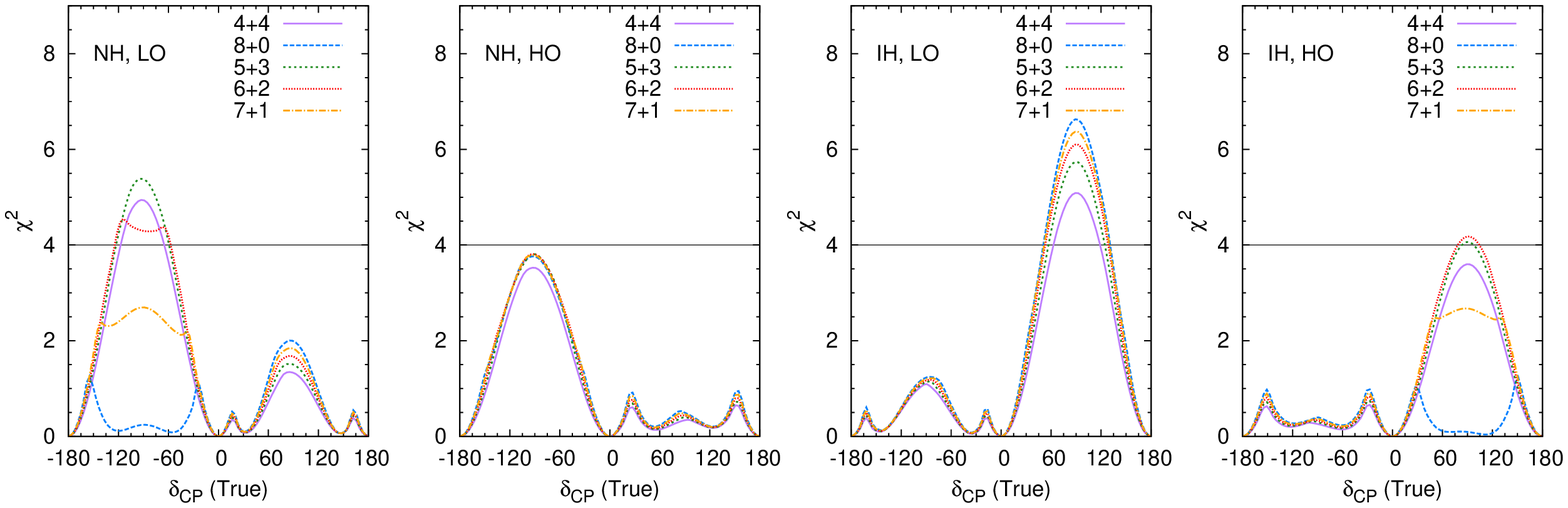}
   \caption{$\delta_{CP}$ discovery potential of T2K for various combinations 
   of neutrino+anti-neutrino runs (in units of $10^{21}$ pot)}
   \label{discovery}
   \end{figure*}
In Fig. ~\ref{cpnurun} we have plotted the CP sensitivity of T2K for the true combinations of NH-LO, NH-HO, IH-LO and IH-HO respectively. For each true combination
we have fitted $\delta_{CP}$ for all four possible combinations of test hierarchy and test octant. 
In these panels we have considered pure neutrino runs of T2K and fixed the true value of $\delta_{CP}$ at $-90^\circ$.
From the top left panel of Fig. ~\ref{cpnurun}, we notice that if NH-LO is the true combination, then a clear best-fit around $\delta_{CP}=-90^\circ$
is not possible. In this case we obtain three best-fits around $-90^\circ$, $0^\circ$ and $135^\circ$. The best-fit of $\delta_{CP}=0^\circ$, and $ 135^\circ$ arise due to the
fake wrong octant solutions corresponding to test IH-HO and NH-HO. Comparing all the four panels of Fig. ~\ref{cpnurun}, we observe that NH-HO is the only true
combination for which an unambiguous clear hint at $\delta_{CP}=-90^\circ$ is possible. This is because for NH-HO, $\delta_{CP} = -90^\circ$ correspond to the maximum point
in the neutrino probability. This high value can not be matched by any other combination of hierarchy, octant and $\delta_{CP}$ 
to generate any degenerate solution. Thus, from these arguments we understand that a hint for $\delta_{CP}=-90^\circ$ by T2K neutrino data would signify normal mass 
hierarchy and higher
octant of $\theta_{23}$ by elimination of the other options.

\subsection{Impact of anti-neutrino run}
\label{subsec:3b}
In Fig. ~\ref{cpantinurun} we have plotted the same figures as that of Fig. ~\ref{cpnurun} but now we are considering equal neutrino and anti-neutrino runs for T2K.
These panels depict the impact of anti-neutrinos run in CP measurement of T2K. 
Comparing with Fig. ~\ref{cpnurun}, we see that inclusion of anti-neutrinos disfavour the solutions that
  appear with the wrong octant. But they do not have any effect on the solutions that come with the wrong hierarchy. In this case we also see that
 the precision, with which other $\delta_{CP}$ values except the true value can be disfavoured, 
 reduces as compared to the full neutrino run because of less statistics. The important conclusion that we draw from this figures is that in this case both NH-LO and
 NH-HO are capable of giving a clear hint of $\delta_{CP}=-90^\circ$.
 
 \subsection{Discovery of CP violation}
\label{subsec:3c}
CP violation (CPV) discovery potential of an experiment is defined by its capability 
of differentiating a true value of $\delta_{CP}$ from the CP conserving values $0^\circ$ and $180^\circ$. 
To see the effect of anti-neutrinos in CPV discovery, in Fig. ~\ref{discovery} we present the CPV $\chi^2$ for different 
combinations of true hierarchy and octant by
dividing the total exposure into different combinations of neutrino and anti-neutrinos in units of $10^{21}$ POT. From Fig. ~\ref{discovery} we see that
for NH-HO and IH-LO, pure neutrino runs of T2K give the best result and addition of anti-neutrino worsen the sensitivity. On the other hand for NH-LO and IH-HO, 
sensitivity is worst for pure neutrino and when anti-neutrinos are added, sensitivity becomes better. For NH-LO the best sensitivity comes for 5+3 and
for IH-HO the best sensitivity comes for 6+2. Here it is interesting to note that for both the cases further addition of anti-neutrino data decreases the CP sensitivity.
From these observations it is quite clear that, the role of the anti-neutrinos is therefore to resolve the octant degeneracy where it is present. Once the $\chi^2$
minima shifts to the right octant, the act of adding anti-neutrino does not help any more, rather it reduces the sensitivity due to less statistics.

\section{Conclusion}
\label{sec:4}
In this work we have studied the possibility of determining the neutrino mass hierarchy and octant of $\theta_{23}$ if further runs of T2K 
confirm the hint of $\delta_{CP}=-90^\circ$. We have also studied the impact
of anti-neutrinos in measuring $\delta_{CP}$. We show that if the recent T2K hint of $\delta_{CP}=-90^\circ$ is established by the pure neutrino runs of T2K, 
then in that case
the true hierarchy and true octant are indicated as NH and HO. But on the other hand if T2K plans to run in equal neutrino and anti-neutrino mode, then a clear
best fit of $\delta_{CP}=-90^\circ$ requires the true hierarchy to be NH. But in this case both LO and HO will be allowed. While analyzing the role of anti-neutrinos in this context
we find that the main role of anti-neutrino is to remove the wrong octant solution by breaking the octant-$\delta_{CP}$ degeneracy. For the true combination of 
hierarchy and octant where this degeneracy is absent, addition of anti-neutrino over neutrino causes a reduction in the CP sensitivity. 
These results are important for optimizing 
the anti-neutrino runs of T2K.


\begin{thebibliography}{99.}

\bibitem{global}
  D.~V.~Forero, M.~Tortola and J.~W.~F.~Valle,
  arXiv:1405.7540 [hep-ph].

\bibitem{t2krecent}
K.~Abe et al. (T2K Collaboration), 
Phys.Rev.Lett. {\bf 112}, 061802 (2014), arXiv:1311.4750. 


\bibitem{akhmedov}

E.~K.~Akhmedov, R.~Johansson, M.~Lindner, T.~Ohlsson, and 
T.~Schwetz, JHEP {\bf 0404}, 078 (2004), hep-ph/0402175;

\bibitem{degen}
H.~Minakata and H.~Nunokawa, 
JHEP {\bf 10}, 001 (2001), hep-ph/0108085;


\bibitem{lisi} 
  G.~L.~Fogli and E.~Lisi,
  Phys.\ Rev.\ D {\bf 54}, 3667 (1996)
  [hep-ph/9604415].

  
  \bibitem{novat2k}
S.~Prakash, S.~K.~Raut, and S.~U.~Sankar, Phys.Rev.
{\bf D86}, 033012 (2012), arXiv:1201.6485.

\bibitem{uma} 
  S.~K.~Agarwalla, S.~Prakash and S.~U.~Sankar,
  JHEP {\bf 1307}, 131 (2013)
  [arXiv:1301.2574 [hep-ph]];

  
  \bibitem{globes}
P.~Huber, M.~Lindner, and W.~Winter, Comput.Phys.Commun. 
{\bf 167}, 195 (2005), hep-ph/0407333; 

\end{thebibliography}
\end{document}